\documentclass[aps, 
               prb, 
               twocolumn, 
               10pt, 
               nofootinbib,
               superscriptaddress,
               amsfonts, 
               amsmath, 
               amssymb]{revtex4-2}

\usepackage{bm}
\usepackage{siunitx}
\usepackage{microtype}
\usepackage{upgreek}
\usepackage{graphicx}
\usepackage{hyperref}
\usepackage{xcolor}
\hypersetup{
    colorlinks,
    linkcolor={blue!80!black},
    citecolor={blue!80!black},
    urlcolor={blue!80!black}
}

\newcommand{\tr}[0]{\mathrm{tr}}
\newcommand{\dd}[0]{\mathrm{d}}

\begin{document}
\title{Percolative supercurrent in superconductor-ferromagnetic insulator bilayers}

\author{A. Maiani}
\altaffiliation{These authors contributed equally to this work.}
\affiliation{Center for Quantum Devices, Niels Bohr Institute, University of Copenhagen, DK-2100 Copenhagen, Denmark}
\affiliation{Nordita, KTH Royal Institute of Technology and Stockholm University,
Hannes Alfvéns väg 12, SE-10691 Stockholm, Sweden}

\author{A. C. C. Drachmann}
\altaffiliation{These authors contributed equally to this work.}
\affiliation{Center for Quantum Devices, Niels Bohr Institute, University of Copenhagen, DK-2100 Copenhagen, Denmark}

\author{L. Galletti}
\affiliation{Center for Quantum Devices, Niels Bohr Institute, University of Copenhagen, DK-2100 Copenhagen, Denmark}

\author{C.~Schrade}
\affiliation{Hearne Institute of Theoretical Physics, Department of Physics \& Astronomy, Louisiana State University, Baton Rouge LA 70803, USA}

\author{Y.~Liu}
\affiliation{Center for Quantum Devices, Niels Bohr Institute, University of Copenhagen, DK-2100 Copenhagen, Denmark}

\author{R. Seoane Souto}
\affiliation{Center for Quantum Devices, Niels Bohr Institute, University of Copenhagen, DK-2100 Copenhagen, Denmark}
\affiliation{Division of Solid State Physics and NanoLund, Lund University, S-22100 Lund, Sweden}
\affiliation{Instituto de Ciencia de Materiales de Madrid (ICMM), Consejo Superior de Investigaciones Científicas (CSIC),
Sor Juana Inés de la Cruz 3, 28049 Madrid, Spain}

\author{S. Vaitiek\.enas}
\affiliation{Center for Quantum Devices, Niels Bohr Institute, University of Copenhagen, DK-2100 Copenhagen, Denmark}

\date{\today}

\begin{abstract}
We report tunneling spectroscopy and transport measurements in superconducting Al and ferromagnetic-insulator EuS bilayers.
The samples display remanent spin-splitting, roughly half the superconducting gap, and supercurrent transport above the average paramagnetic limit. 
We interpret this behavior as arising from the interplay between two characteristic length scales: the superconducting coherence length, $\xi$, and the magnetic domain size, $d$. 
By comparing experimental results to a theoretical model, we find $\xi/d \approx 10$.
In this regime, spin-averaging across the micromagnetic configuration can locally suppress superconductivity, resulting in percolative supercurrent flow.
\end{abstract}
\maketitle

\section{Introduction}
The coexistence of superconductivity and ferromagnetism typically leads to intriguing ground states, emerging from 
the competing interplay of ferromagnetic spin alignment and superconducting singlet pairing~\cite{Meservey1994, Buzdin2005}.
Predicted manifestations include proximity-induced spin-splitting~\cite{Meservey_PRB_1975, Xiong_PRL_2011, Hubler_PRL_2012}, domain wall superconductivity~\cite{Yang_NatMat_2004}, spatial modulation of the order parameter~\cite{Fulde_PR_64, Larkin_1964, Izyumov_PhysUsp_2002}, and the onset of unconventional superconducting pairing states~\cite{Bergeret_RMP_2005, Seleznyov2023}.
These phenomena play a key role in designing current-phase relations in hybrid Josephson junctions~\cite{Sellier_PRL_2004, Strambini_EPL_2015, Razmadze_2023_Supercurrent, Maiani_PRB_2023, Birge_APL_2024}, spin-polarized supercurrents for spintronics~\cite{Eschrig_RepProgPhys_2015}, superconducting diodes~\cite{Ando_Nature_2020, Strambini_NC_2022, Hou_PRL_2023}, and thermoelectric devices~\cite{Machon_PRL_2013, Kolenda_PRL_2016, Bergeret_RMP_2018}.
Moreover, magnetic textures, such as skyrmions, domain walls, or helical spin states, in proximity to superconductors can lead to unique electron pair correlations~\cite{Linder_PRB_2009, Nakosai_PRB_2013, Dahir_PRB_2020, Spuri_PRR_2024, Sardinero_PRB_2024, Plastovets_PRB_2024}.
Controlling the interplay between these orders enables the synthesis of spin-orbit interactions~\cite{Klinovaja_PRL_2012, Kjaergaard_PRB_2012, Desjardins_NM_2019}, spin-triplet pairing~\cite{Linder_RMP_2019, Silaev_PRB_2020, Diesch2018}, and topological superconductivity~\cite{Sau_PRL_2010, Schrade_PRL_2015, Vaitiek_NatPys_2020, Maiani_PRB_2020, Khindanov_PRB_2021, Escribano_PRB_2021, Liu_PRB_2021, Liu_PRB_2022, Escribano_NPJQM_2022}, while also providing a platform to explore the role of percolation phenomena in various phase transitions~\cite{Porat_PRB_2015, Azpeitia_MaterAdv_2021}.

In pristine superconductor--ferromagnetic-insulator heterostructures, interfacial scattering can induce strong exchange splitting, $h$, with little contribution from stray fields~\cite{Tokuyasu_PRB_1988, Li2013, Liu_NL_2019}.
For homogeneous samples at low temperatures, the superconducting order parameter, $\Delta$, depends weakly on $h$ but gets fully suppressed at the paramagnetic limit, $h_{\rm C}$~\cite{Heikkila_PSS_2019}.
At zero temperature, the transition to the normal state in conventional superconductors is first order, occurring at a critical splitting $h_{\rm C0} = \Delta / \sqrt{2}$, known as the Chandrasekhar-Clogston limit~\cite{Chandrasekhar_APL_1962, Clogston_PRL_1962}.
As the temperature increases, the critical field decreases, and the transition becomes second-order~\cite{Sarma_JPCS_1963}.

At a given point within the superconductor, the induced $h$ is proportional to the local micromagnetic domain configuration, $m(\bm{r})$, averaged over a characteristic spin-averaging length, $\xi$~\cite{Matthias_PRL_1960, Champel_PRB_2005b}.
At low energies, $\xi$ is dominated by superconducting correlations and corresponds to the superconducting coherence length~\cite{Tokuyasu_PRB_1988, Bergeret_PRB_2004}.
Under non-equilibrium conditions, $\xi$ is renormalized due to quasiparticle diffusion~\cite{Morten_PRB_2004, Hijano_PRB_2022}.
For samples with an elementary magnetic domains (magnetic grains) of size $d\ll \xi$, the ferromagnetic proximity effect weakens superconductivity through spin-flip scattering~\cite{Bruno_PRB_1973}, while the induced exchange splitting is approximately homogeneous and approaches the sample-average $\langle h \rangle$~\cite{Ivanov_PRB_2006, Aikebaier_PRB_2019}.
For $d\gg\xi$, the superconductivity can be suppressed locally but persists at the domain boundaries~\cite{Aladyshkin_PRB_2003, Jing_SST_2014}, as has been observed, for instance, in heterostructures with out-of-plane magnetization~\cite{Yang_NatMat_2004, Iavarone_NatCom_2014, Stellhorn_NJP_2020}.
Several previous experiments on superconductor--ferromagnetic-insulator bilayers reported measurements compatible with the $d\gg\xi$ case~\cite{Strambini_PRM_2017, Rouco_PRB_2019b, Hijano_PRR_2021, Machon_2022}.

Here, we investigate superconducting Al and the ferromagnetic-insulator EuS bilayers using two complementary experimental techniques.
Tunneling spectroscopy reveals hysteretic superconductivity characterized by a remanent $h\approx130~\mu$eV at zero field, nearly half the superconducting gap $\Delta = 280~\mu$eV, and negligible spin-splitting at the coercive field.
This highlights the importance of the spin-averaging effects in our samples.
Resistance measurements on proximitized Al bars show $h_{\rm C}$ that increases with bar width, eventually surpassing the paramagnetic limit.
We develop a minimal, fittable model that captures the essential physics of the system by combining the Usadel formalism for the superconductor with a simplified microscopic model for the magnet.
Our model is self-contained, with all input parameters measured experimentally.
By fitting the model, we estimate $\xi/d \approx 10$, contrasting with previous experiments on similar bilayers~\cite{Strambini_PRM_2017, Rouco_PRB_2019b, Hijano_PRR_2021, Machon_2022}.
The comparable magnitudes of the competing length scales imply that the statistical fluctuations in magnetic domain configuration play an important role in determining the local superconducting properties in the material.
A complementary resistive-network model suggests this behavior stems from supercurrent percolation around normal regions in the sample.

\begin{figure}[t!]
\includegraphics[width=\linewidth]{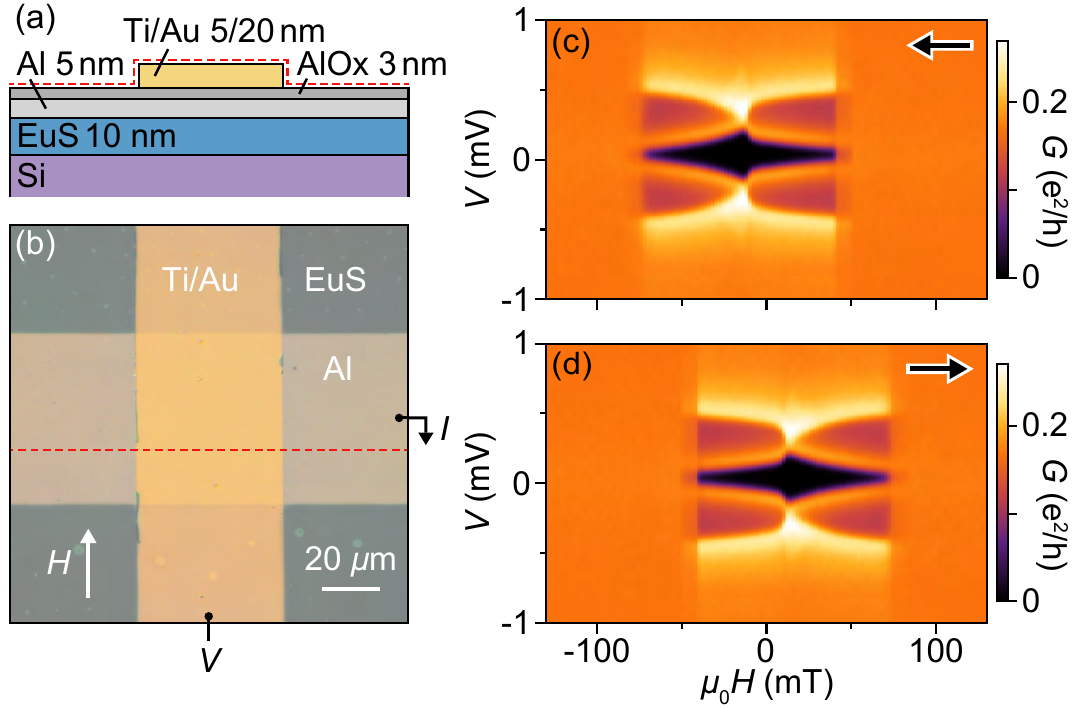}
\caption{\label{fig:1}
(a) Schematic cross-section of Al-EuS bilayer with a Ti/Au lead on top of native AlOx layer, used for tunneling spectroscopy.
(b) Optical micrograph of $60\times50~\mu$m$^2$ junction with measurement setup.
(c) and (d) Differential conductance, $G$, as a function of voltage bias, $V$, and in-plane magnetic field, $H$, measured for the junction shown in (b), sweeping $H$ from (b) positive to negative and (c) negative to positive.
The spectra display a superconducting gap with sweep-direction-dependent evolution of the coherence peak splitting.
The data were taken after polarizing the sample at $\mu_0 H_\parallel = \pm 200$~mT.
}
\end{figure}


\section{Experimental Setup}

Measurements were performed on an Al/EuS (5/10 nm) bilayer grown on an insulating Si substrate.
The Al bars were lithographically defined using selective wet etch.
Tunneling junctions were formed by metalizing strips of Ti/Au (3/20 nm) across selected Al bars, using the native AlOx (3~nm) as the barrier; see Figs.~\ref{fig:1}(a) and \ref{fig:1}(b).
Further details on sample preparation are given in Appendix~\ref{app:growth}.
Two samples with several devices each showed consistent results.
In the main text, we report representative data from Sample 1.
Supporting data from other devices are summarized in Appendix~\ref{app:additional_measurements}.

Transport measurements were performed using standard ac lock-in techniques in a dilution refrigerator with a three-axis (1,~1,~6)~T vector magnet and a base electron temperature of 40~mK.
The dc lines used for addressing the devices were equipped with RF and RC filters, adding an additional line resistance of $R_{\rm line} = 6.7~$k$\Omega$.
Two-terminal differential conductance measurements of the tunneling junctions were performed using $10~\mu$V ac-voltage excitation at frequencies between 105 and 113~Hz.
Four-terminal differential resistance measurements of the Al bars were carried out using a 10~nA ac-current excitation at frequencies between 13 and 44~Hz.

\section{Measurements}
We begin with tunneling spectroscopy of the junctions.
Differential conductance, $G = \dd I/\dd V$, measured for a $60\times50~\mu$m$^2$ junction as a function of voltage bias, $V$, shows a hysteretic evolution with an in-plane magnetic field, $H$ [Figs.~\ref{fig:1}(c) and \ref{fig:1}(d)].
Starting at $\mu_0 H = 200$~mT, the junction displays a featureless $G(V)$.
Decreasing the field toward zero, the conductance spectrum becomes gapped at $\mu_0 H \approx 50$~mT, showing split peaks positioned symmetrically around $V=0$.
The splitting decreases as the field passes through zero, becoming negligible around $\mu_0 H = -15$~mT, where the spectral gap is maximal.
The peaks gradually split again for more negative $H$ and the spectrum turns featureless at $\mu_0 H \approx 80$~mT.
An inverse behavior is observed when sweeping from negative to positive $H$.

\begin{figure}[t!]
\includegraphics[width=\linewidth]{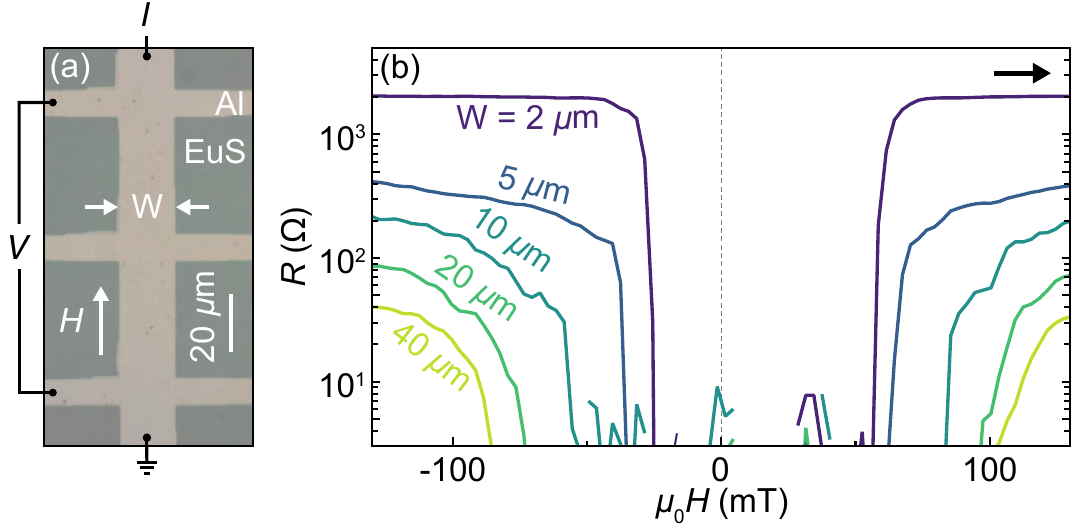}
\caption{\label{fig:2}
(a) Optical micrograph of 20~$\mu$m wide Al bar showing the four-probe measurement setup.
(b) Differential resistance, $R$, as a function of in-plane magnetic field, $H$, measured for Al bars of different widths, showing a superconducting window centered away from $H=0$.
The onset of the resistive state increases with bar width.
}
\end{figure}

The measured tunneling spectra resemble the superconducting density of states with spin-split coherence peaks.
The nonlinear and hysteretic response to the external field can be attributed to the proximity-induced exchange coupling.
The separation between the coherence peaks provides a measure of the sample-average exchange splitting.
In our sample, the peaks merge around $\mu_0 H = \pm 15$~mT, around the coercive field of thin-film EuS~\cite{Muller2011, Strambini_PRM_2017, Hijano_PRR_2021}, implying that a typical magnetic domain is smaller than the spin-averaging length.
The transition to a normal state at higher magnetic fields can be understood by domain magnetization, yielding an average induced splitting larger than the paramagnetic limit, $\Delta/\mu_{\rm B} \sqrt{2} \approx 3.4$~T, with $\Delta = 280~\mu$eV measured at the coercive field and Bohr magneton $\mu_{\rm B}$.
We note that the transition is gradual, with an intermediate state showing smeared features for roughly $10$~mT after the onset and before the end of the gapped spectrum, extending between $\mu_0 H \approx \pm(40$--$50)$ and $\mp(70$--$80)$~mT, depending on the sweep direction [see Figs.~\ref{fig:1}(c) and \ref{fig:1}(d)]. This suggests that superconductivity can survive locally in regions where the induced exchange splitting is below the paramagnetic limit.

We next investigate supercurrent transport in proximitized Al bars [Fig.~\ref{fig:2}(a)].
Four-terminal differential resistance, $R=\dd V/\dd I$, measured at zero dc current bias as a function of $H$ for Al bars of varying widths from 2 to 40~$\mu$m reveals a hysteretic window of suppressed resistance; see Fig.~\ref{fig:2}(b).
The size of the window displays a characteristic dependence on the bar width, spanning $\approx 100$~mT around the coercive field, ranging from $\mu_0 H\approx -40$ to $70$~mT for the narrower bars, similar to the gapped region identified in the junctions.
In contrast, for the wider bars, the window broadens, extending to nearly 200~mT, from $\mu_0 H\approx -90$ to $110$~mT for the 40~$\mu$m Al bar, with the onset of the resistive state exceeding the paramagnetic limit measured from the tunneling spectroscopy.
We note that the transition to the normal state is gradual for all widths but becomes progressively smoother for wider bars.

The observed behavior can be interpreted as the magnetization-driven superconductor-normal metal transition.
Resistance remains suppressed as long as there is a superconducting path along the bar, allowing for current to flow without dissipation.
With increasing $H$, the flipping of elementary domains can form macroscopic clusters much larger than $d$, where uniform magnetization locally suppresses superconductivity. This leads to the interruption of the percolative supercurrent paths by normal regions, resulting in finite $R$.
In this regime, $R$ depends on the length of the normal regions the current passes.
For larger $H$, the magnetization drives the whole sample normal, and $R$ saturates; see Fig.~\ref{fig:2}(b).
At a given $H$, the supercurrent percolation is more likely to be interrupted in a narrow bar than in a wider one.


\section{Theoretical model}
To support our interpretation, we introduce a theoretical framework that uses experimentally measured parameters to model supercurrent transport in bars of different widths.
As a soft ferromagnet, EuS displays easy-plane magnetization in thin-film geometry due to shape anisotropy~\cite{GomezPerez_2020}. 
Therefore, we model the ferromagnetic insulator as a two-dimensional grid of elementary domains $m(\bm{r})=\pm 1$, each of size $d$, that are constrained to align either parallel or antiparallel to the external magnetic field.
The effective local exchange field is described as 
\begin{equation}
    \label{eq:exchange_field}
   h(\bm{r}) = h_\mathrm{S} \langle m \rangle_{\xi} + E_{\rm Z}\,.
\end{equation}
The first term is the proximity-induced exchange field, where $h_\mathrm{S}$ is the saturation exchange splitting that defines the strength of the proximity effect, and $\langle m \rangle_\xi$ is a local average magnetization with range $\xi$.
The second term is the Zeeman energy, $E_{\rm Z} = g \mu_B \mu_0 H / 2$, with electron g factor $g = 2$, Bohr magneton $\mu_\mathrm{B}$, and external magnetic field $\mu_0 H$.
We stress that the local average exchange field $h(\bm{r})$ is an inhomogeneous quantity representing the induced exchange splitting spatial distribution in contrast with the global average $\langle h \rangle$ that takes a single value for a given domain configuration.

\begin{figure}[t!]
\includegraphics[width=\linewidth]{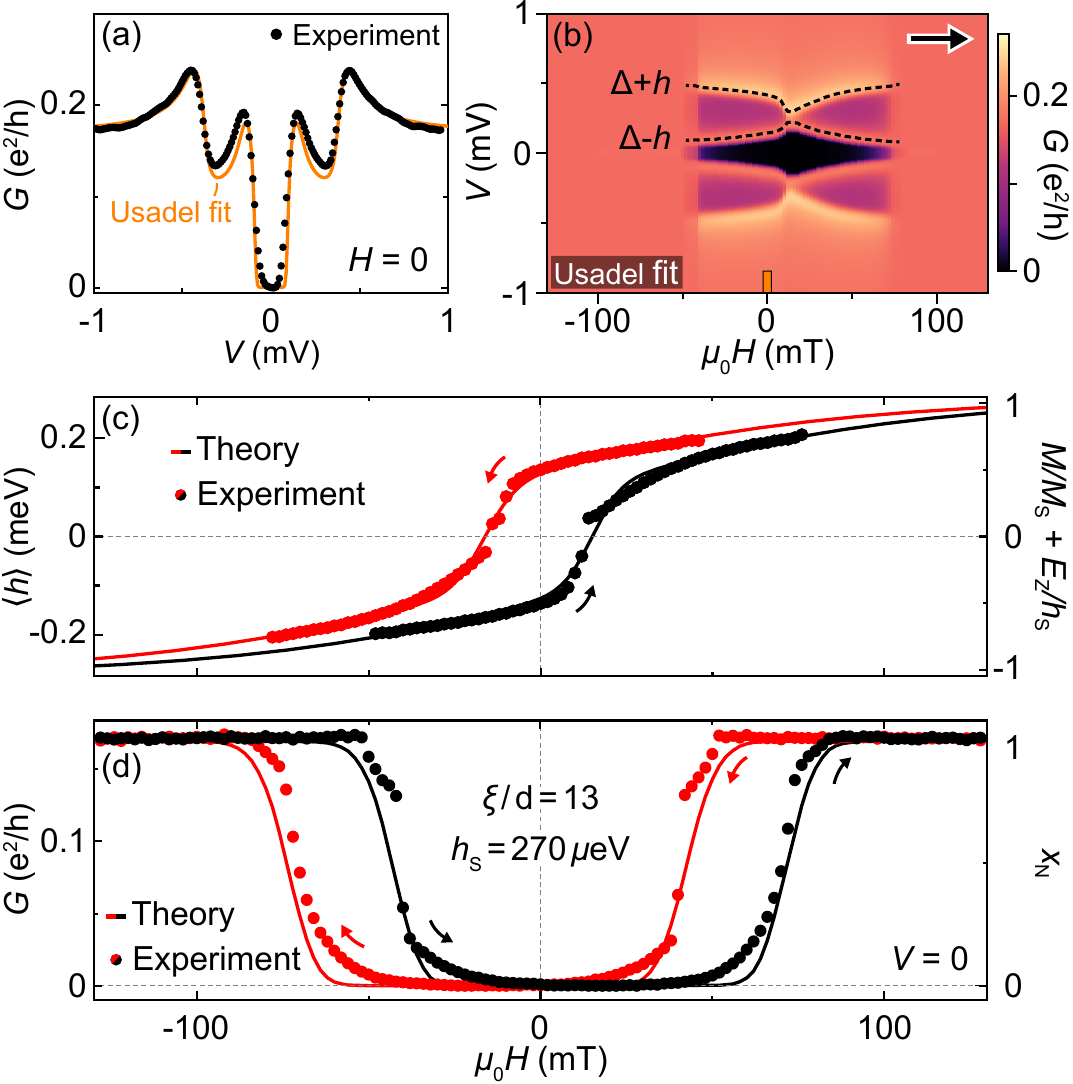}
\caption{\label{fig:3}
(a) Differential conductance, $G$, as a function of voltage bias, $V$, measured for $60\times50~\mu$m$^2$ junction at zero in-plane magnetic field, $H=0$, (data points) and Usadel model fit (solid curve).
(b) Usadel model fit applied to the data in Fig.~\ref{fig:2}(d). The extracted mean superconducting gap, $\Delta=280~\mu$eV.
(c) Extracted sample-averaged exchange field, $\langle h \rangle$, (data points) and extrapolated magnetization curve, $M$, (solid curves) normalized to its saturation value, $M_{\rm S}$. 
(d)~Comparison of zero-bias conductance measured as a function of $H$ (data points) with the fraction of the sample in the normal state, $x_{\rm N}$, (solid curves). 
The theory curves in (c) and (d) are calculated using a saturation exchange splitting of \mbox{$h_{\rm S}=270~\mu$eV} and a ratio between superconducting coherence length and magnetic domain size, $\xi/d=13$. 
}\end{figure}

\begin{figure*}[t!]
\includegraphics[width=\linewidth]{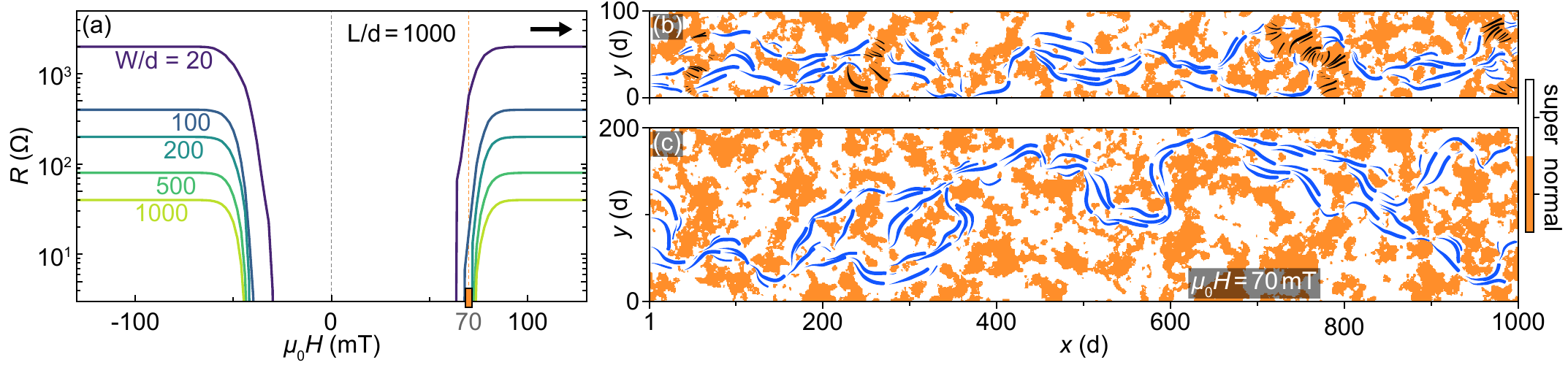}
\caption{\label{fig:4}
(a)~Calculated resistance, $R$, as a function of the applied magnetic field, $H$, for grids of $W \times L$ domains. The onset of $R$ increases with $W$.
(b)~Simulated distribution of superconducting (white) and normal (orange) regions for a grid with $W/d = 100$ and $L/d=1000$ at $\mu_0 H_\parallel=\SI{70}{mT}$, overlaid with a stream plot that illustrates the paths of supercurrent (blue) and normal current (black).
Note that several normal regions span the entire width of the grid, disrupting the supercurrent flow.
(c)~Similar to (b) but for $W/d=200$, showing uninterrupted supercurrent percolation along the grid.
}
\end{figure*}

We extract the global average exchange splitting, $\langle | h | \rangle$, and the superconducting gap, $\Delta$, by numerically fitting the tunneling spectroscopy data [Figs.~\ref{fig:1}(c) and \ref{fig:1}(d)] to the nonlinear Usadel model~\cite{Usadel_PRL_1970}, following the procedure summarize in Appendix~\ref{app:usadel_details}.

The conductance can be expressed as
\begin{equation}
    \label{eq:conductance}
     G(V) = G_\mathrm{N} [1 - x_\mathrm{N}]\,n_\mathrm{S}(e V) + G_\mathrm{N} x_\mathrm{N}\,,
\end{equation}
where $G_\mathrm{N}$ is the normal state differential conductance, $x_\mathrm{N} \simeq G(V=0) / G_\mathrm{N}$ is the fraction of the sample in the normal state, and $n_\mathrm{S}$ is the thermally broadened superconducting density of states described by the homogeneous Usadel equations that include average exchange splitting~\cite{Bergeret_RMP_2005, Bergeret_RMP_2018, Heikkila_PSS_2019}. The output of the Usadel fit procedure is summarized in Figs.~\ref{fig:3}(a) and \ref{fig:3}(b). 
A more detailed description of the fitting procedure and results is given in Appendix~\ref{app:fit}.

In the second step, we simultaneously estimate $h_\mathrm{S}$ and the $\xi/d$ ratio through an iterative procedure. The sample average magnetization, $M(H)$, is obtained by fitting the normalized extracted exchange splitting, $\langle h \rangle / h_\mathrm{S}$, after accounting for the Zeeman splitting, to a double hyperbolic tangent function~\cite{Meszaros_JPCS_2011}
\begin{equation}\label{eq:magnetization}
    M(H) = M_\mathrm{S} \frac{\sum_{i=1, 2} C_i \tanh\left[N_i (H \pm H_\mathrm{C})\right]}{\sum_{i=1, 2} C_i},
\end{equation}
where $M_\mathrm{S}$ is the saturation magnetization defined as $M/M_\mathrm{S} =\langle m \rangle$, whereas $C_i$, $N_i$, and $H_\mathrm{C}$ are the amplitude, scaling, and coercive-field fit parameters, respectively; see solid curves in Fig.~\ref{fig:3}(c).
The magnetization curve is then used to generate a randomized evolution of the micromagnetic configuration $m(\bm{r}; H)$, whose average reproduces the extracted $M(H)/M_\mathrm{S}$. We assume that superconductivity is suppressed at points where the local exchange field, $|h|$, calculated using Eq.~\eqref{eq:exchange_field}, exceeds the Chandrasekhar-Clogston limit, $h_\mathrm{C0}=\Delta/\sqrt{2} = (198\pm3)~\mu$eV, with average $\Delta$ taken from the fit in Fig.~\ref{fig:3}(a) and (b).
With this, the expected $x_\mathrm{N}$ can be calculated using the average number of normal regions and compared to the experimental zero-bias differential conductance; see Fig.~\ref{fig:3}(d).
The procedure is repeated until the optimal set of $h_\mathrm{S}$ and $\xi/d$ is found by minimizing the error in $x_\mathrm{N}$, resulting in $h_\mathrm{S}=(270\pm10)~\mu$eV and $\xi/d = 13\pm3$.
Within our framework, the main uncertainties in estimating the $\xi/d$ ratio arise from the fit error of $h_{\rm C}$ and $h_{\rm S}$, though both values are reasonable and consistent with the literature~\cite{Hijano_PRR_2021}.
A more detailed description of the model is given in Appendix~\ref{app:model} and Ref.~\cite{Zenodo}.

Finally, we use a resistive network model~\cite{Strelniker_PRB2007} to simulate the supercurrent transport in proximitized Al bars.
We follow the same procedure as in the previous step to generate $m(\bm{r}; H)$ for a grid of $W\times L$ domains.
We assume strong coupling between the superconducting regions and, therefore, neglect the phase dynamics that may be relevant for very narrow superconducting constrictions~\cite{Kresin_PR_2006}. For each value of $H$, the resistance of the bar can be calculated by assigning a finite resistivity $\rho_\mathrm{N} = 40~\Omega\square$, taken from Fig.~\ref{fig:2}(b), to the normal regions and $\rho_\mathrm{S} = 0$ to the superconducting regions and solving the continuity equation for the current, $\bm{\nabla} \cdot \bm{j} = \bm{\nabla} \cdot \rho^{-1} \bm{\nabla} \varphi = 0$, where $\bm{j}$ is the current density and $\varphi$ is the electric potential. We apply a unit of voltage bias across neighboring domains and compute the resulting current flow to calculate the resistance.
In the normal state, the bar resistance saturates at $\rho_\mathrm{N}\,L / W$.
The simulated resistance for grids with a fixed $L = 1000\,d$ displays a hysteretic superconducting window that increases with $W$, qualitatively reproducing the experimental observations; see Fig.~\ref{fig:4}(a).

To better understand this behavior, we compare the distribution of superconducting and normal regions at the onset of the resistive state, simulated for two bars with $W = 100\,d$ and $200\,d$; see Figs.~\ref{fig:4}(b) and \ref{fig:4}(c).
At $\mu_0 H = 70$~mT, the narrower bar displays several segments where the normal regions span the entire width [see orange patches in Fig.~\ref{fig:4}(b)], interrupting supercurrent flow along the bar and causing dissipation.
At the same magnetic field, the wider bar exhibits a comparable density of normal regions but none that extends across the full width.
In this case, the supercurrent can percolate around the normal regions, allowing for dissipationless current flow. 
We note that while the precise magnetic field value for the onset of the resistive state is inherently random, it strongly depends on both $W$ and $L$.
For a fixed length, narrower bars are more susceptible to supercurrent interruptions due to the presence of normal regions.

Our resistive network model disregards the $\Delta$ dependence on $h$, which may lead to an increase in $\xi$ close to the paramagnetic limit. 
Furthermore, while the simulated grids maintain the aspect ratios of the experimental Al bars, they are scaled down because of computational constraints.
These simplifications could explain the quantitative differences between the experimental [Fig.~\ref{fig:2}(b)] and simulated [Fig.~\ref{fig:4}(a)] resistance traces.

\section{Conclusions}
In summary, we have studied the coexistence of superconductivity and ferromagnetism in Al-EuS bilayers. 
By measuring tunneling spectroscopy and supercurrent transport in the proximitized Al, we observed persistent supercurrent percolation above the proximity-induced spin splitting reaching the paramagnetic limit.
We argue that this behavior arises from the interplay between two characteristic length scales: the superconducting coherence length, $\xi$, and the minimal magnetic domain size, $d$. 
To support this interpretation, we introduce a minimal, fittable model, suggesting that $\xi\gtrsim d$.
In this previously unexplored regimen, where the two length scales are comparable, statistical fluctuations in the induced magnetization create characteristic superconducting patterns with locally suppressed superconductivity.
Our findings establish a flexible framework for investigating how magnetic textures affect superconductivity by using parameters extracted from tunneling conductance measurements to fit a micromagnetic model.
This approach provides insights into phase transitions in disordered magnets and potential applications for hybrid superconducting devices. These results highlight the promise of magnetic insulator components for future studies on synthetic quantum phases.

\section*{Acknowledgments}
We thank Karsten Flensberg, Charles Marcus, Egor Babaev, and Filipp Rybakov for valuable discussions, Peter Krogstrup and Claus S\o rensen for contributions to materials growth.
We acknowledge support from the Danish National Research Foundation, the Danish Council for Independent Research \textbar Natural Sciences, European Research Council (Grant Agreement No. 856526), European Innovation Council (Grant Agreement No. 101115548), Spanish CM “Talento Program” (project No. 2022-T1/IND-24070), Spanish Ministry of Science, Innovation, and Universities through Grant PID2022-140552NA-I00, NanoLund, a research grant (Projects No. 43951 and No. 53097) from VILLUM FONDEN, and funding from the Wallenberg Initiative on Networks and Quantum Information (WINQ).


\appendix

\section{SAMPLE PREPARATION}
\label{app:growth}

\begin{figure}[t!]
    \centering
    \includegraphics[width=\linewidth]{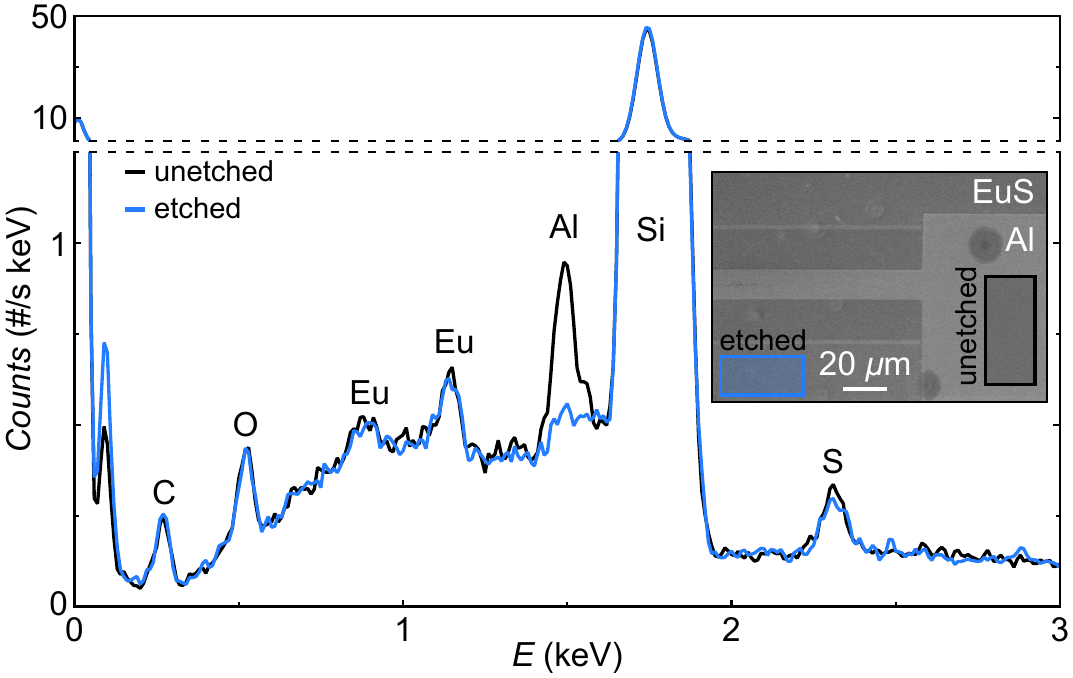}
    \caption{Comparison of two energy dispersive spectra taken for the Al/EuS bilayer, integrated over two areas with etched and unetched Al. The main difference between the two spectra is the suppressed Al peak for the etched area, whereas the Eu and S peaks remain nearly unchanged.
    Inset: Scanning electron micrograph of a test chip after Al etching. Exposed EuS (etched regions) appear darker, whereas the unetched Al appears brighter.}
    \label{fig:sfig1}
\end{figure}
The Al-EuS bilayer studied in this work was prepared on a 2-inch intrinsic (100) FZ Si wafer with resistivity $>10^4\,\Omega$\,cm (Sil'tronix Silicon Technologies).
First, the silicon wafer was etched in a $5\%$ HF solution for 10\,s to strip the native oxide.
After cleaning, the wafer was loaded into a dedicated metal deposition chamber that is part of a molecular beam epitaxy (MBE) system with a background pressure $<10^{-10}$~Torr.
Both layers were deposited using electron beam evaporation.
The 10~nm EuS layer was grown at room temperature with an average growth rate of 0.01~nm/s.
Next, the sample was annealed \textit{in situ} for one hour at $400^\circ$C, followed by 6~hours of cooling.
The 7~nm Al layer was evaporated at a rate of 0.06~nm/s with a nominal substrate temperature of $-156^\circ$C, as measured by a thermo-coupling back sensor.
Calibration using a test wafer showed the surface temperature of roughly $-20^\circ$C.
Finally, the surface of the Al layer was oxidized \textit{in situ} for 10~s using 10~Torr partial oxygen pressure, converting the top 2~nm of Al into a uniform layer of approximately $3$~nm AlOx.


\begin{figure}[t!]
    \centering
    \includegraphics[width=\linewidth]{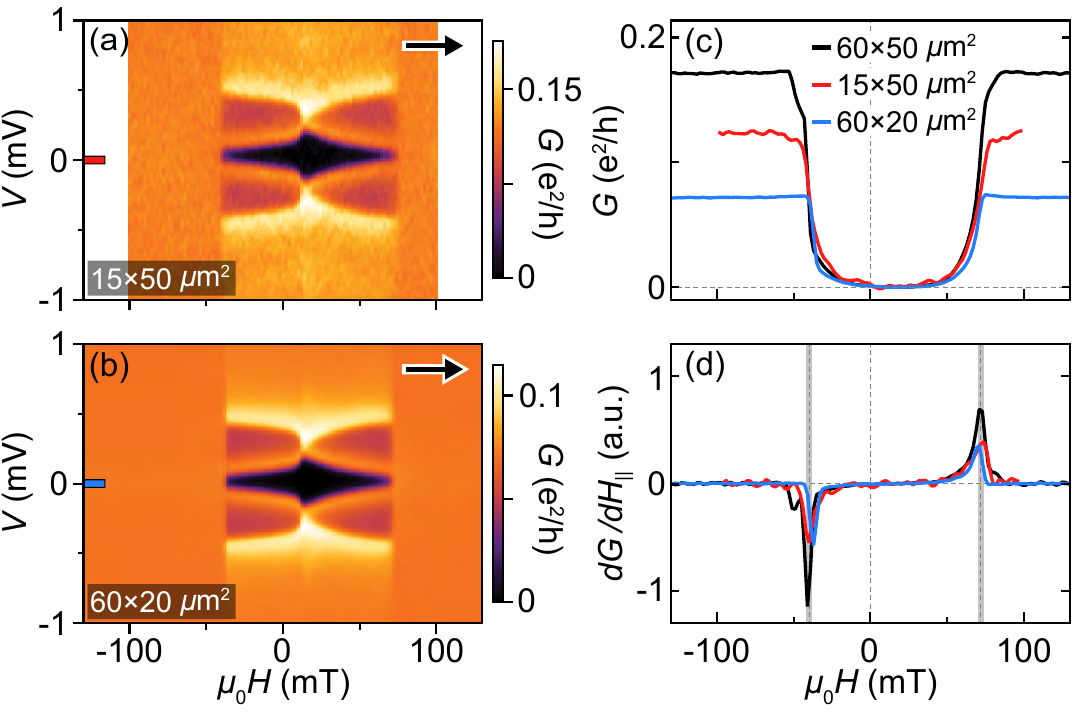}
    \caption{(a)~Differential conductance, $G$, as a function of voltage bias, $V$, and in-plane magnetic field, $H$, for $15\times50~\mu$m$^2$ tunneling junction on Sample 1.
    Sweep direction is indicated by the arrow.
    (b)~Same as (a) but for $60\times20~\mu$m$^2$ junction.
    (c)~Zero-bias $G$ line-cuts taken from (a), (b), and main-text Fig.~1(d).
    (d) Same as (c) but numerically differentiated with respect to $H$.
     Black dashed lines indicate transitions from normal to superconducting regime at $\mu_0 H = -(39 \pm 1)$~mT and back to normal at $\mu_0 H = (71 \pm 1)$~mT. The gray bands represent errors from standard deviation across the junctions.}
    \label{fig:sfig2}
\end{figure}

\begin{figure*}[t!]
    \centering
    \includegraphics[width=\linewidth]{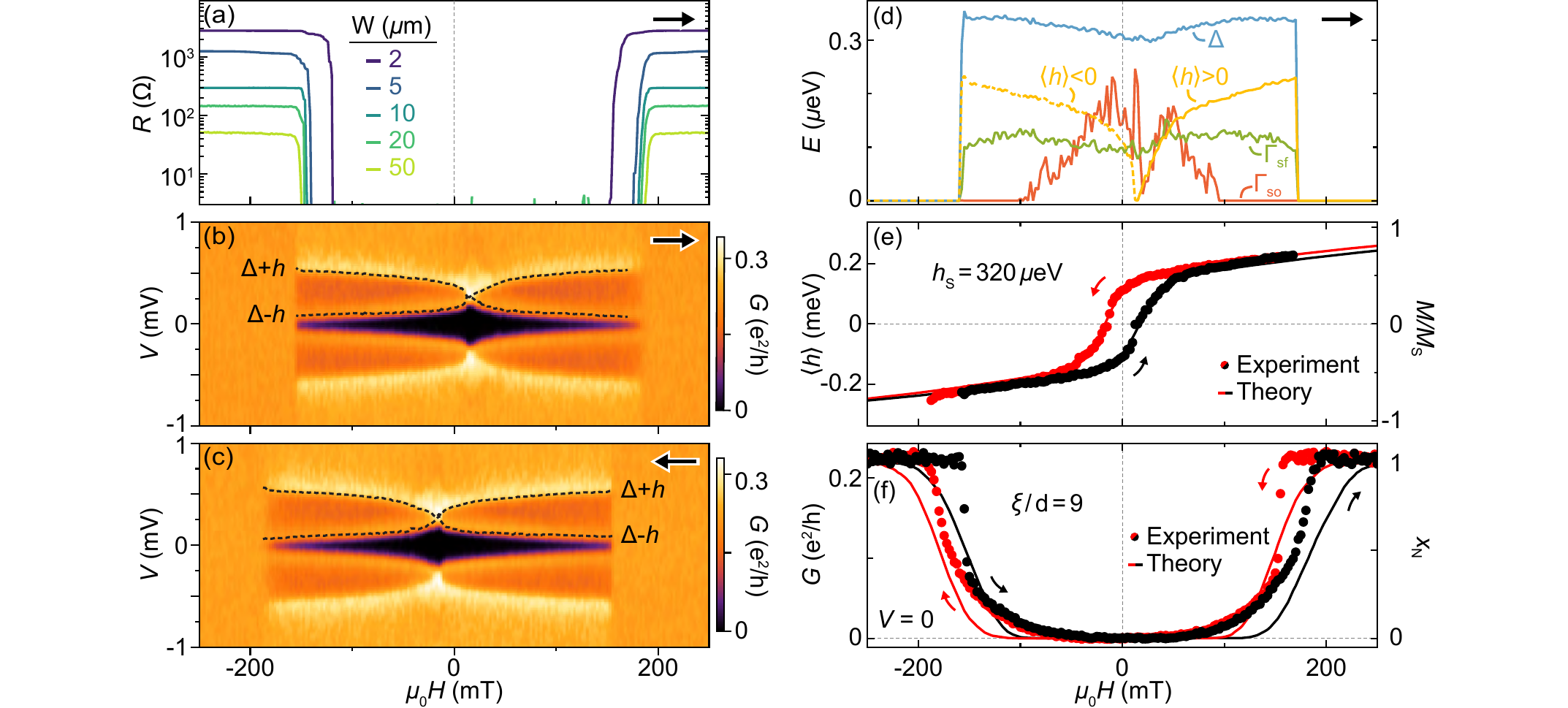}
    \caption{(a) Differential resistance, $R$, as a function of in-plane magnetic field, $H$, for Al bars of different widths on Sample~2.
    Sweep direction is indicated by the arrow. Data for 2, 5, and 50~$\mu$m bars were taken in a three-terminal configuration. Background resistance was subtracted using a five-point average around $H=0$.
    (b) Differential conductance, $G$, as a function of voltage bias, $V$, and $H$ for $25\times50~\mu$m$^2$ tunneling junction on Sample~2.
    (c) Same as (b) but for opposite field-sweep direction.
    (d) Output of the Usadel fitting routine performed on data in (b), giving superconducting gap, $\Delta$, effective exchange splitting, $h$, spin-flip relaxation rate, $\Gamma_\mathrm{sf}$, and spin-orbit scattering rate, $\Gamma_\mathrm{so}$.
    (e) and (f) Results of the percolative model.
    (e) Average exchange splitting, $\langle h \rangle$, (left axis) and fitted magnetization normalized to its saturation value, $M/M_{\rm S}$, (right axis) as a function of $\mu_0 H$. The fit yields paramagnetic limit $h_{\rm C} = 230 = \Delta/\sqrt{2} = (232\pm7)~\mu$eV.
    (f) Zero-bias conductance, $G$, (left axis) from (b) and (c), and deduced normal fraction of the sample, $x_{\rm N}$, (right axis) as a function of $H$. The fit yields saturation spin-splitting $h_{\rm S} = 320~\mu$eV and coherence-to-domain size ratio, $\xi/d = 9$.
    }
    \label{fig:sfig3}
\end{figure*}

\begin{figure}[t!]
    \centering
    \includegraphics[width=\linewidth]{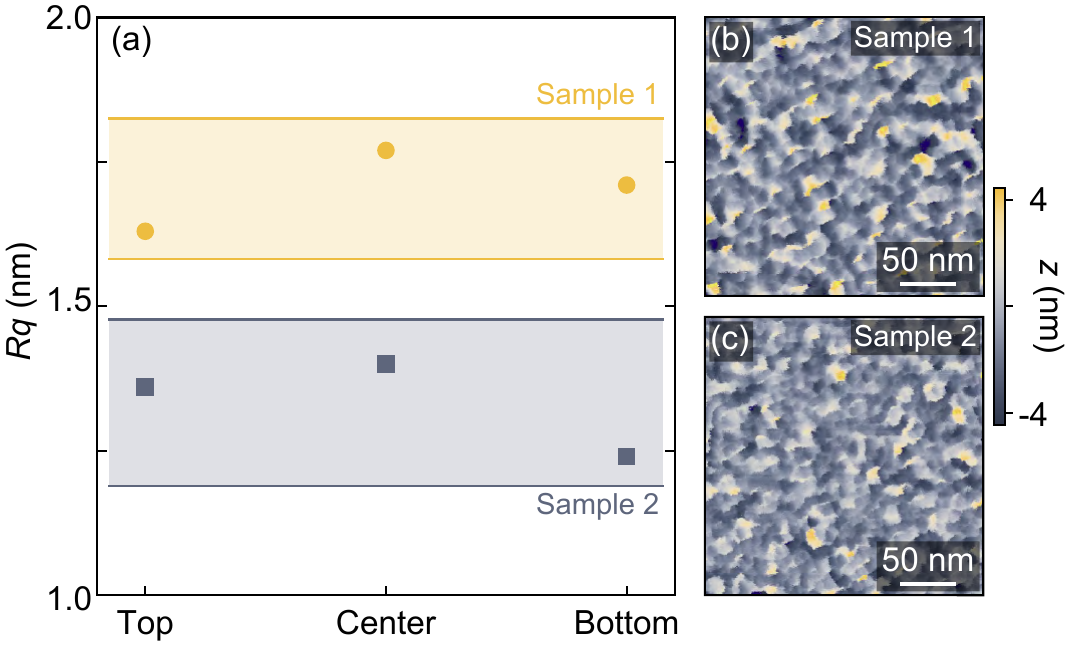}
    \caption{(a) Surface roughness, $Rq$, measured on the exposed EuS surface at three different locations for both samples. The bands illustrate six-fold standard errors centered around the mean values, indicating statistically significant differences between the surface roughness of the two samples.
    (b)~Atomic force micrograph taken at the bottom of Sample 1, with $Rq = 1.7$~nm.
    (c) Same as (b) but taken at the center of Sample 2, with $Rq = 1.4$~nm.}
    \label{fig:sfig4}
\end{figure}

For processing, the wafer was diced into $5\times5$~mm$^2$ pieces.
The devices were fabricated using standard electron beam lithography techniques.
Al was selectively etched using a combination of AR 300-80 (new) adhesion promoter, CSAR 62 (AR-P 6200) 4$\%$ resist, and IPA:TMAH 0.2 N developer 19:1 solution with 6 min etching time at room temperature.
Energy dispersive spectroscopy confirmed selective Al etching; see Fig~\ref{fig:sfig1}.
Normal metal Ti/Au (3/20~nm) leads were metalized using another layer of CSAR 62 (AR-P 6200) 4$\%$ resist.

\section{ADDITIONAL MEASUREMENTS}
\label{app:additional_measurements}

Four samples were investigated, each with several tunneling junctions and Al bars.
The majority of the studied devices showed qualitatively similar behavior.
Outliers were not considered, such as junctions with spurious spectral features hinting at inhomogeneous barriers and Al bars with inconsistent normal state resistance likely due to incomplete etching.
The main findings are presented in the main text using representative data from Sample~1.
Additional tunneling spectroscopy data from Sample~1 are summarized in Fig.~\ref{fig:sfig2}.
Fourteen junctions, with areas ranging from 120~$\mu$m$^2$ to 6000~$\mu$m$^2$, showed transition from normal to superconducting regime at $\mu_0 H = -(39 \pm 1)$~mT, and back to normal at $\mu_0 H = (71 \pm 1)$~mT, when sweeping from negative to positive $H$ [Fig.~\ref{fig:sfig2}(d)]. Here, the errors indicate the standard deviation across the investigated junctions.

Sample 2, nominally identical to Sample 1, showed qualitatively similar behavior. It had a slightly higher switching field to and from the superconducting regime and a narrower field range for percolative supercurrent, see Fig.~\ref{fig:sfig3}(a-c).
Fitting of the Usadel model to the tunneling spectra suggests that both the superconducting gap, $\Delta$, and spin-flipping strength, $\Gamma_\mathrm{sf}$, are approximately field-independent [Fig.~\ref{fig:sfig3}(d)], consistent with literature~\cite{Hijano_PRR_2021}.
The effective exchange spin splitting, $h$, displays a minimum at the coercive field around $\pm 15$~mT, suggesting a switch in the average magnetization direction. 
The coercive field of thin-film EuS depends on growth conditions~\cite{Miao2009} and varies with temperature, including below the superconducting transition temperature of Al~\cite{Li2013}.
The deduced spin-orbit strength, $\Gamma_\mathrm{so}$, displays a lobe-shaped field dependence, with a sharp decrease around the coercive field, where accurate estimation becomes challenging due to the merging of the coherence peaks.
We speculate 
The deduced paramagnetic limit $h_\mathrm{C0}=\Delta/\sqrt{2} = (232\pm7)~\mu$eV, with average $\Delta$ taken from the fit in Fig.~\ref{fig:sfig3}(d).
We speculate that the larger gap observed in Sample 2 compared to Sample 1 results from a slight increase in the native oxide over time, reducing Al thickness, which in turn increases the superconducting gap and decreases the coherence length~\cite{Court2008}.
Fitting of the percolative model to the extracted average exchange splitting, $\langle h \rangle$, and zero-bias conductance yields $h_{\rm S}= (320\pm60)~\mu$eV, higher than for Sample~1, and $\xi/d = 9\pm7$, slightly smaller than for Sample~1 [Fig.~\ref{fig:sfig3}(e) and (f)].
The errors were estimated by propagating the uncertainty in $h_\mathrm{C}$ in the model, giving a conservative upper bound.
The magnetization curves deduced from the up and down $H$ sweeps show residual paramagnetic behavior at high magnetic fields after they meet  [Fig.~\ref{fig:sfig3}(e)].

\begin{figure*}[t!]
    \centering
    \includegraphics[width=\linewidth]{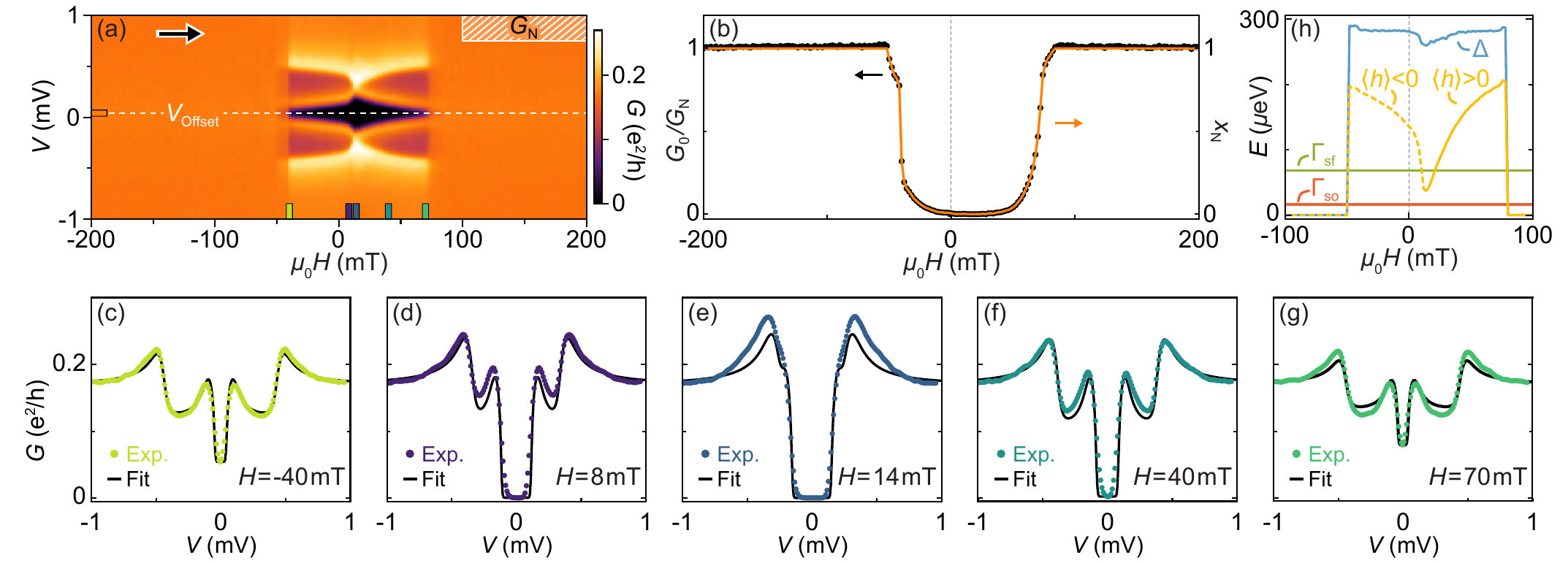}
    \caption{Illustration of the Usadel fitting algorithm steps applied to the data shown in the main-text Fig.~1(d).
    (a) First, we correct the voltage-bias offset, $V_{\rm Offset}$, and extract the normal state conductance, $G_{\rm N}$, by taking the average of a region at high field and bias.
    (b) Next, we extract the fraction of the sample in the normal state, $x_\mathrm{N}$, defined as $G(V=0)/G_\mathrm{N}$, constrained to range between 0 and 1.
    (c)-(g) Finally, we fit the measured conductance with a Usadel model by a minimization routine. 
    (h) As a result of the fitting routine, we extract the Usadel model parameters: pairing potential $\Delta$, average exchange splitting $\langle | h | \rangle$, spin-flip relaxation rate $\Gamma_\mathrm{sf}$, and spin-orbit scattering rate $\Gamma_\mathrm{so}$.}
    \label{fig:sfig5}
\end{figure*}

Atomic force microscopy (AFM) of the exposed EuS surfaces reveals significant differences in the topography of the two samples, see Fig.~\ref{fig:sfig4}.
The measured surface roughness, $Rq$, is $(1.70\pm0.04)$~nm for Sample 1, and $(1.33\pm0.05)$~nm for Sample 2, with the uncertainties given by the standard errors from independent measurements at different locations.
We speculate that the different surface roughness leads to different sizes of elementary magnetic domains.
We tentatively ascribe the variations in surface topography between the two nominally identical samples to small thermal gradients across the wafer during the growth.

\section{USADEL FIT PROCEDURE}
\label{app:fit}

Differential conductance of a normal metal--insulator--superconductor (NIS) tunneling junction is given by~\cite{Giaever_1961_Study}
\begin{equation}
    \frac{\mathrm{d} I}{\mathrm{d} V} = \frac{G_\mathrm{N}}{n_\mathrm{N}} \int_{-\infty}^{+\infty} n( \omega) \left[ - \partial_\omega f( \omega - e V) \right] \mathrm{d}\omega,
\end{equation}
where $G_\mathrm{N}$ is the normal state conductance, $e$ is the electron charge, $n$ is the density of states in Al, $n_\mathrm{N} = n(\omega \gg \Delta)$ is the normal density of states, $f(\omega, T)$ is the Fermi-Dirac distribution with temperature $T$ and energy $\omega$, and $V$ is the bias voltage.
Here and for the rest of the discussion, we set $\hbar=1$.

We assume that, in general, a fraction $x_\mathrm{N}$ of the sample is in the normal state, while the rest remains in the superconducting phase with a pairing potential unaffected by orbital magnetism.
In this case, the effective density of states in the Al film can be expressed as 
\begin{equation}
     n( \omega) = (1 - x_\mathrm{N}) n_\mathrm{S}(\omega) + x_\mathrm{N} n_\mathrm{N},
\end{equation}
where $n_\mathrm{S}(\omega)$ is the density of states of a homogeneous superconductor.

We model the spin-split superconductor with the Usadel equation~\cite{Usadel_PRL_1970}, which has been extended for ferromagnet-superconductor heterostructures~\cite{Bergeret_RMP_2005, Bergeret_RMP_2018, Heikkila_PSS_2019}.
In the Usadel formalism, $n_\mathrm{S}(\omega)$ can be defined using four parameters:
pairing potential $\Delta$, average exchange splitting affecting the surviving superconducting regions $\langle | h | \rangle$, spin-flip relaxation rate due to magnetic impurities $\Gamma_\mathrm{sf}= 3/2\,\tau_\mathrm{sf}^{-1}$, and spin-orbit scattering rate $\Gamma_\mathrm{so}= 3/2 \,\tau_\mathrm{so}^{-1}$, where $\tau_\mathrm{sf}$ and $\tau_\mathrm{so}$ are the spin-flip and spin-orbit scattering times.
In general, $\Delta$ and $\langle | h | \rangle$ determine the position of the spin-split coherence peaks, $\Gamma_\mathrm{sf}$ broadens the peaks, while $\Gamma_\mathrm{so}$ affects the height difference between the peaks for the two spin components.

\subsection*{Fitting algorithm}
The model is highly non-linear and thus requires a physically motivated fitting algorithm.
We start by correcting any small voltage offset, defined by the middle of the superconducting gap, and by fixing the temperature to the base electron temperature of the dilution refrigerator, $T=40$~mK~\cite{Nichele2017}.
We then fit the remaining parameters using the following steps. 
\begin{enumerate}
    \item \textbf{Normal state conductance} $G_\mathrm{N}$.\\
    The normal state conductance, $G_\mathrm{N}$, is estimated by taking the average of the differential conductance at high voltage bias and large magnetic field, far above the Chandrasekar-Clogston limit [Fig.~\ref{fig:sfig5}(a)]. 
    \item \textbf{Normal fraction} $x_\mathrm{N}$.\\
    We estimate the normal fraction of the sample as $x_\mathrm{N} \simeq G(V=0)/G_\mathrm{N}$, assuming that the subgap tails of the coherence peaks due to thermal broadening are negligible [Fig.~\ref{fig:sfig5}(b)]. 
    We note that because of the uncertainty in $G_\mathrm{N}$, the estimated $x_\mathrm{N}$ can be $\gtrsim 1$. Another reason $G$ can exceed $G_\mathrm{N}$ can be attributed to the effect of quantum fluctuations above the paramagnetic limit~\cite{Khodas_PRL_2012}. Therefore, we constrain the values to range between 0 and 1, [Fig.~\ref{fig:sfig5}(c)].    
    \item \textbf{Usadel parameters}.\\
    For the remaining Usadel parameter, we explore two different fitting approaches. 
    
    For Sample 1, we first fit a single differential conductance trance at $\mu_0 H = 8$~mT, which we use to fix $\Gamma_\mathrm{sf}$ and $\Gamma_\mathrm{so}$. We then fit the traces at other field values for $\Delta$ and $\langle | h| \rangle$. The results of this approach are summarized in Fig.~\ref{fig:sfig5}(d).
    
    For Sample 2, we run a free optimization routine, starting with the zero-field trace, followed by the fitting of $H > 0$ and then $H < 0$ data.
    The output from the previous fit is used as the initial guess for the next field value.
    The results of this approach are summarized in Fig.~\ref{fig:sfig3}(d).

    The two alternatives were chosen to ensure robustness in parameter estimation, particularly in scenarios where one of the two routines was prone to convergence issues. Cross-validation showed that the results were consistent within acceptable error margins, reinforcing the reliability of using different approaches.
    
\end{enumerate}

\section{MICROSCOPIC MODEL}
\label{app:model}

\subsection*{Magnetic insulator}
To better understand the observed experimental behavior, we introduce a simple microscopic model using the following experimentally-motivated considerations 
\begin{enumerate}
    \item The ferromagnet is composed of \textit{elementary domains} that behave as single spins.
    The size of these domains is approximately constant and equal to $d$. They are distributed homogeneously enough to be described as a grid.
    \item The proximity-induced magnetization is governed primarily by surface exchange interactions, allowing the system to be described as two-dimensional. Any stray fields arising from surface roughness are considered negligible.
    \item The elementary domains can point only in two anti-collinear directions, defined by the external magnetic field $H$, such that the micromagnetic configuration is $m(\bm{r})=\pm 1$ where $\bm{r}$ is the position of a domain.
    \item In our model there is no spatial correlation between the magnetic grain orientations. We note that the spatial correlation is reintroduced in the superconductor via the local averaging.
\end{enumerate}

To describe the magnetic domain flipping, we consider a static, non-interacting picture and describe the probability of a domain pointing in the positive direction with respect to the sign of the magnetization to be given by 
\begin{equation}
    p = \frac{1}{2}\Big(1 + \frac{M}{M_\mathrm{S}}\Big),
\end{equation}
where the ratio of total to saturation magnetization $M/M_\mathrm{S}$ ranges between -1 and 1.
This can be thought of as a simplified version of a more general model discussed in Ref.~\cite{Glauber1963}, where the magnetization state determines the flipping probabilities.
For a magnetic field sweep, we can generate a series of micromagnetic configurations, $m(\bm{r}; H)$. For each simulation run, we initialize the system in the homogeneously polarized state, $m(\bm{r}; H \to -\infty) = -1$.
At each increment of the external magnetic field, a number of domains are flipped until the average value $\langle m \rangle = M(H)/M_\mathrm{S}$ matches the extracted data from the Usadel fitting as described below.
While our model incorporates stochastic domain flipping, which accounts for natural variability, it also includes simplifications, such as constant domain size and negligible spatial correlation in the magnet. Despite these approximations, the model effectively captures the qualitative behavior of supercurrent transport.

\subsection*{Superconductor}
At the interface with the superconductor, the magnet generates a proximity-induced \textit{surface exchange field}, $h_\mathrm{S} m(\bm{r})$.
For completeness, we include the Zeeman splitting induced by the external field, $E_\mathrm{Z} = {g \mu_B \mu_0 H}/{2}$, but note that its effect is negligible.
The response of a superconductor to this strong, non-uniform exchange field is highly nonlinear and nonlocal.
To reproduce this complex superconducting response, we introduce a simple model that combines two operations---a nonlocal linear filter and a local nonlinear step transition---described below.

First, we define the \textit{effective exchange field} 
\begin{equation}
    h = h_\mathrm{S} \langle m \rangle_\xi + E_\mathrm{Z}   
\end{equation}
where $h_\mathrm{S} \langle m \rangle_{\xi} = h_\mathrm{S} m(\bm{r}) \otimes K_\xi(\bm{r})$ represents the effective proximity-induced exchange field at point $\bm{r}$. This is obtained via a weighted moving average defined by the convolution filter
\begin{equation}
    K_\xi (\mathbf{r}) = \frac{1}{2 \pi \xi^2} K_0(r/\xi)\,,
    \label{eq:filter}
\end{equation}
where $K_0$ is the modified Bessel function of the second kind and has an effective radius of $\xi$.
This definition of $K_\xi$ is motivated by the fact that both the spin-diffusion in the normal state and the linear response theory derived below are described by screened Poisson equations with fundamental solution in two dimensions given by Eq.~\eqref{eq:filter}. In the numerical implementation, the divergence of the filter at the origin is regularized by integrating numerically over the lattice.

Second, we introduce a local nonlinear operation to describe a first-order transition to the normal state at a given point within the superconductor if the local exchange splitting $|h|$ exceeds the paramagnetic limit $h_\mathrm{C}$; otherwise, it remains superconducting. Mathematically, we can describe the normal metallic regions as
\begin{equation}
X_n = \{ \mathbf{r} : | h | > h_\mathrm{C} \},   
\end{equation}
which defines the theoretically expected normal fraction $\tilde{x}_\mathrm{N} = |X_n| / LW$, where $L$ is the length and $W$ is the width of the simulated grid.

\begin{figure}[t!]
    \centering
    \includegraphics[width=\linewidth]{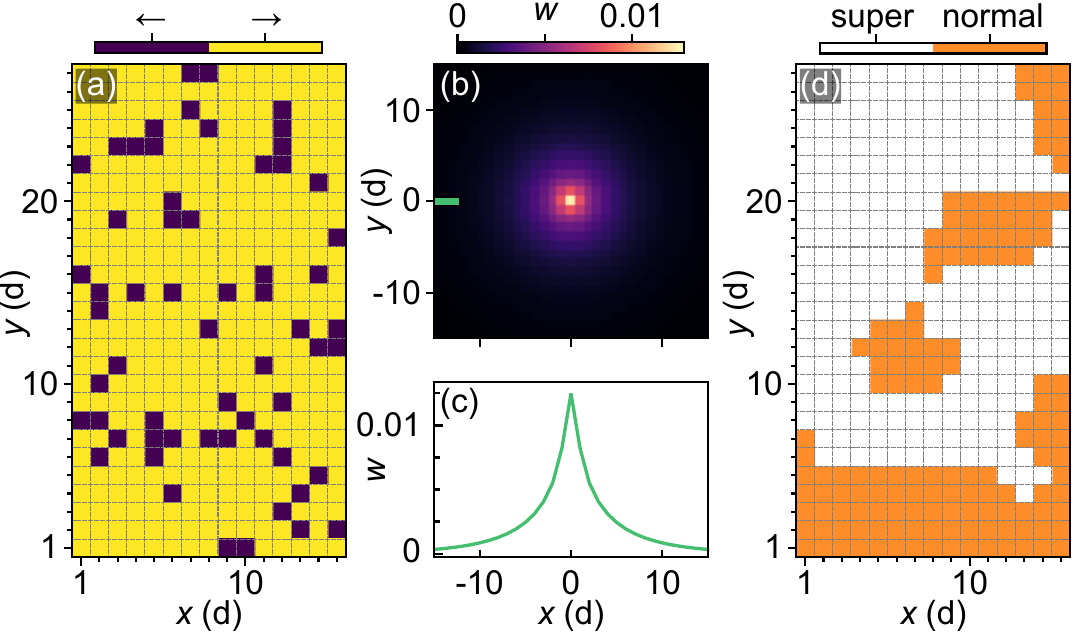}
    \caption{(a) Zoom-in micromagnetic domain configurations of the grid from a $200~d\times1000~d$ realization at in-plane magnetic field $\mu_0 H = 70$~mT.
    (b) and (c) Averaging filter weight, $w$, calculated using Eq.~\eqref{eq:filter} with $\xi/d=13$.
    (d)~Spatial distribution of the superconducting (white) and normal (orange) regions, calculated by applying the averaging filter to the micromagnetic configuration in (a). The grid turns normal where local average spin-splitting exceeds paramagnetic limit $h > h_\mathrm{C}$.}
    \label{fig:sfig6}
\end{figure}

The effect of the two operations, namely the local averaging of micromagnetic configuration and the nonlinear transition to the normal state, is summarized in Fig.~\ref{fig:sfig6}. It is apparent that for a given micromagnetic domain configuration, the sample turns normal in regions with one domain species but remains superconducting in regions with mixed domains.

Note that in the strong exchange field limit, where the local exchange field can reach values close to or higher than the paramagnetic limit, the superconducting behavior is nonlinear, as the spin-averaging length depends on the pairing potential. Therefore, local averaging through a convolution filter can only serve as a qualitative representation of the physics of the structure.

\subsection*{Theoretical results of the percolation model}
Before applying the model to the experimental data, we consider the possible zero-bias conductance curves predicted by the model.

In the limit of $g=0$, there is no transition to the normal state if $h_\mathrm{S}<h_\mathrm{C}$, irrespective of the other parameters.
The normal state regions appear for $h_\mathrm{S}>h_\mathrm{C}$.
In this case, the relationship between the spin-averaging length and the domain size determines the behavior of the system.
For $\xi/d\ll1$, there is effectively no averaging, and the sample is always in the normal state, with a potential exception of the domain-wall superconductivity.
For $\xi/d\sim1$, the finite local averaging results in a smooth transition.
For $\xi/d\gg1$, the transition to the normal state is sharp due to effective global averaging.

When $g\neq0$, the Zeeman effect enables a local phase transition to the normal state even for $h_\mathrm{C}>h_\mathrm{S}$.
We can quantify the importance of the Zeeman field by considering the critical field 
\begin{equation}
    H^* = \frac{h_\mathrm{C0} - h_\mathrm{S}}{g \mu_B \mu_0/2}\,.
\end{equation} 
Below \(H^*\), the whole sample is superconducting as the total exchange splitting is not sufficient to suppress superconductivity at any point.
For $H > H^*$, the total exchange splitting can locally suppress superconductivity.  

In this work, $H^*$ is always negative due to $h_\mathrm{S}>h_\mathrm{C}$, suggesting that the Zeeman field plays a minor role.
For the analysis, we fix $g=2$ as we do not expect a strong renormalization of the electronic $g$ factor in Al.

\subsection*{Model fitting}
To apply the model to the experimental data, we first extract the magnetization curve $M(H)$ by fitting the main-text Eq.~(3) to the measured sample average spin-splitting extracted from the Usadel fitting. In the frame of the microscopic model, we can interpret the quantity extracted from the Usadel fitting as $\langle | h | \rangle$.

Given the small sample size, where $\xi \gtrsim d$, the effective field $h$ shows statistical fluctuations. We assume that, while these fluctuations are present, they are not large enough to cause random changes in the sign of $h$ throughout the sample. Under this assumption, we can approximate
\begin{equation}
     \langle | h | \rangle \approx | \langle  h \rangle | =  \Big| h_\mathrm{S} \langle m \rangle + E_\mathrm{Z} \Big|\,.
\end{equation}

\noindent With this, one gets
\begin{equation}
    h_\mathrm{S} \frac{M}{M_\mathrm{S}} = h_\mathrm{S} \langle {m} \rangle \approx \pm \langle | h | \rangle \pm E_\mathrm{Z}\,,
\end{equation}
allowing for fitting the sample-averaged magnetization $M(H)$ to the extracted $\langle | h | \rangle$, after adjusting the sign change due to the domain flipping around the coercive field and detrending for the external field Zeeman contribution.

\begin{figure}[t!]
    \centering
    \includegraphics[width=\linewidth]{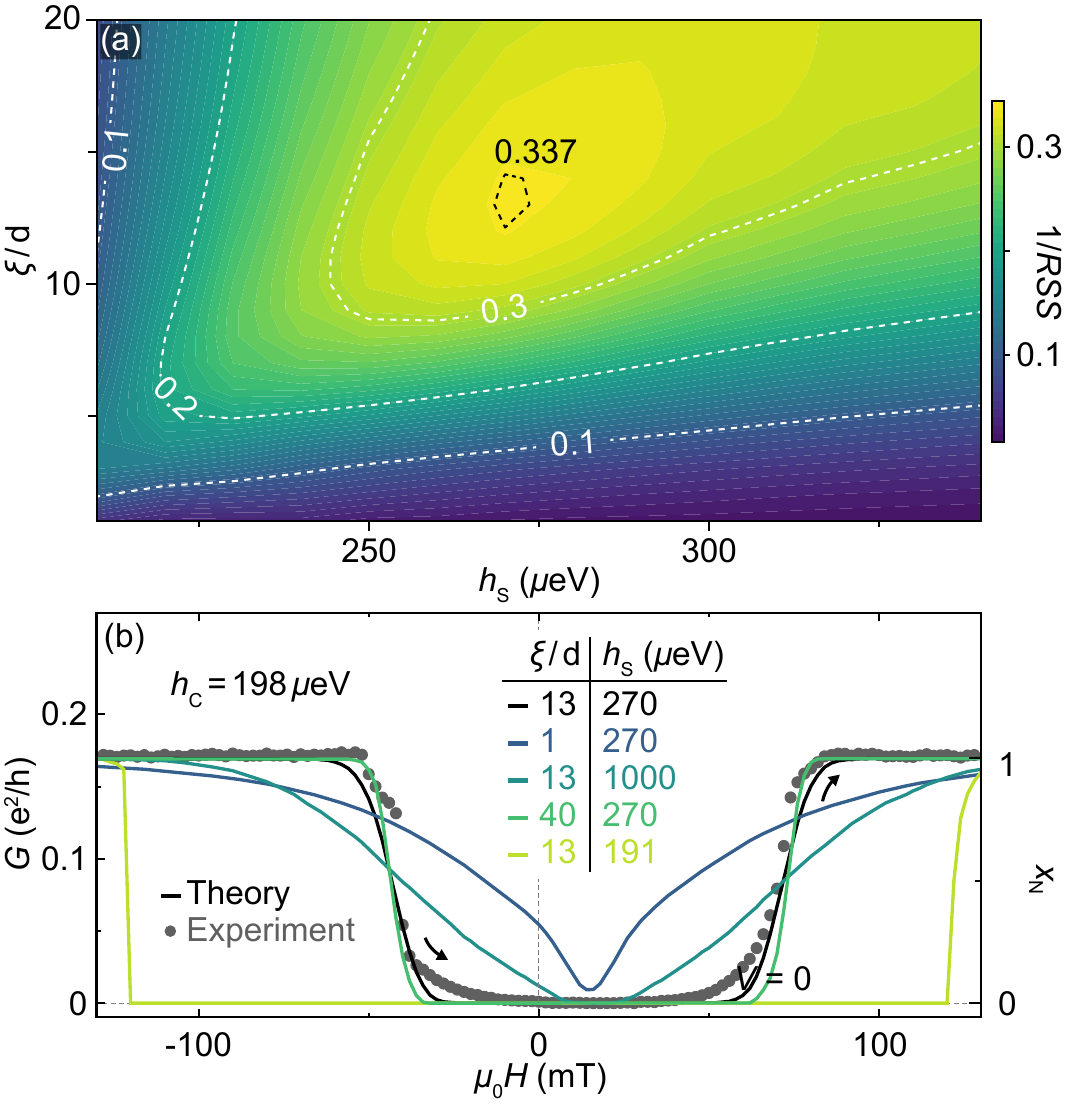}
    \caption{
    (a) Inverse of the residual sum of squares, 1/RSS, as a function of saturation exchange splitting, $h_\mathrm{S}$, and the ratio between the superconducting coherence length and minimal domain size, $\xi/d$.
    The data were calculated using the minimal model for zero-bias conductance to compare the calculated and experimentally determined normal fraction of the sample, $x_\mathrm{N}$, shown in the main-text Fig.~3(d).
    (b) A comparison between the expected $x_\mathrm{N}$ evolutions with in-plane field $H$ calculated for different values of $h_\mathrm{S}$ and $\xi/d$.}
    \label{fig:sfig7}
\end{figure}

The magnetization curve is then used to generate a collection of matching micromagnetic configurations, $m(\bm{r}; H)$, and to calculate an expected $\tilde{x}_\mathrm{N}(\xi, h_s)$ that will depend on the choice of $\xi$ and $h_\mathrm{S}$.
We determine the most reasonable set of $\xi/d$ and $h_\mathrm{S}$ by minimizing the residual sum of squares of the calculated and experimentally determined normal fractions, $\mathrm{RSS} = [\tilde{x}_\mathrm{N}(\xi, h_\mathrm{S}) - x_\mathrm{N}]^2$.
The final fit results are shown in the main-text Figs.~3(c) and 3(d), while the value of $1/\mathrm{RSS}$ is displayed in Fig.~\ref{fig:sfig7}(a).
A comparison between the predicted $\tilde{x}_\mathrm{N}$ evolutions for different values of $h_\mathrm{S}$ and $\xi/d$ is shown in Fig.~\ref{fig:sfig7}(b).

\section{USADEL MODEL DETAILS}
\label{app:usadel_details}

In the time-reversed hole basis, $\psi = (\psi_\uparrow, \psi_\downarrow, -\psi_\downarrow, \psi_\uparrow)$, the Usadel equation for the quasiclassical propagator $\check{g}$ in Matsubara representation reads as
\begin{equation}
    D \nabla\cdot(\check{g}\nabla\check{g}) + [i  \omega_n \tau_3 \sigma_0 - i \boldsymbol{h}\cdot\boldsymbol{\sigma} \tau_3 - \check{\Delta} -\check{\Sigma}, \check{g}] = 0,
\label{eq:usadel}
\end{equation}
where $D$ is the diffusion constant, $\omega_n=2\pi \frac{(2 n +1)}{T}$ with temperature $T$ are the Matsubara frequencies, $\bm{h}$ is the exchange splitting, $\check{\Delta}= \Delta \tau_1$ is the singlet order parameter and $\check{\Sigma}$ are additional self-energies~\cite{Aikebaier_PRB_2019}.
We recall that the diffusion constant is defined as $D = v_{\rm F} \ell/3$, where $v_F$ is the Fermi velocity and $\ell$ is the elastic mean free path, while the coherence length for diffusive superconductor is $\xi = \sqrt{D/\Delta}$~\cite{Bergeret_RMP_2005, Bergeret_RMP_2018, Heikkila_PSS_2019}.
The real-energy equation is obtained by the substitution $\omega_n \to - i \omega$.
Eq.~\eqref{eq:usadel} is complemented with the normalization constraint $\check{g}^2 = 1$.

To account for the spin-flipping scattering and spin-orbit coupling, we include the two spin-imbalance relaxation terms in the self-energy.
The magnetic spin-flip term reads
\begin{equation}
    \Sigma_\mathrm{sf} = \frac{\bm{\sigma}\cdot\tau_3\check{g}\bm{\sigma} \tau_3}{8  \tau_\mathrm{sf}}\,,
\end{equation}
where $\tau_\mathrm{sf}$ is related to normal state spin diffusion length by $\lambda_\mathrm{sf} = \sqrt{\tau_\mathrm{sf} D}$~\cite{Morten_PRB_2004}.
The spin-orbit scattering term reads
\begin{align}
\Sigma_\mathrm{so}&=\frac{\bm{\sigma}\cdot\check{g}\bm{\sigma}}{8   \tau_\mathrm{so}}\,.
\end{align}
After finding a solution for $\check{g}$, the local density of states is calculated as
\begin{equation}
    n_\mathrm{S}( \omega) = n_\mathrm{N} \Re\, \tr [\check{g} \tau_3 \sigma_0],
\end{equation}
where $\Re$ denotes the real part.
\vspace{2pt}

\subsection*{Numerical method}
To solve the Usadel equation, we adopt the hyperbolic $(\theta, \boldsymbol{M})$ parametrization~\cite{Ivanov_PRB_2006}.
In this parametrization, the quasiclassical propagator reads 
\begin{equation}
\begin{split}
    \check{g} = & (\cos\theta M_0  + i \sin\theta \boldsymbol{M}\cdot \boldsymbol{\sigma}) \tau_3 + \\
    & ( \sin\theta M_0 - i \cos\theta \boldsymbol{M}\cdot\boldsymbol{\sigma})\tau_1\,,    
\end{split}
\end{equation}
with $M_0^2 - \mathbf{M}^2 = 1$.
With this, the Usadel equation can be split into a scalar (the $i\tau_2 \sigma_0$ component) and a vector (the $\tau_2 \boldsymbol{\sigma}$ component) coupled partial differential equations
\begin{equation}
\label{eq:usadel_1}
\begin{split}
&D\nabla^2 \theta + 2 M_0(\Delta\cos\theta + i \omega\sin\theta)-\\
- &2 \cos\theta \boldsymbol{h}\cdot\boldsymbol{M} -\frac{(2 M_0^2 + 1) \sin(2 \theta)}{4\tau_\mathrm{sf}}=0, 
\end{split}
\end{equation}

\begin{equation}
\label{eq:usadel_2}
\begin{split}
&D\left(\boldsymbol{M}\nabla^2M_0  - M_0\nabla^2\boldsymbol{M}\right) + 2 \boldsymbol{M} (\Delta \sin\theta - i \omega \cos\theta) -\\
- &2  \sin\theta \boldsymbol{h} M_0 
+ \left[\frac{1}{\tau_\mathrm{so}}+ \frac{\cos(2 \theta)}{2 \tau_\mathrm{sf}}\right] M_0 \bm{M}= 0\,.
\end{split}
\end{equation}


We discretize Eqs.~\eqref{eq:usadel_1} and \eqref{eq:usadel_2} using a finite-difference scheme and solve the nonlinear partial differential equations with the Newton method.
The code for this procedure is provided in Ref.~\cite{pyUsadel}.

\subsection*{Spin-averaging length}
\label{sm:spin_averaging_length}
To gain further insight into the behavior of a superconductor subjected to an inhomogeneous ferromagnetic proximity effect, we simplify the Usadel model for analytical treatment by following the methodology outlined in Ref.~\cite{Hijano_PRB_2022}.
The homogeneous solution of the Usadel equations in the absence of self-energy scattering terms takes the form of
\begin{equation}
\label{eq:Usadel_homogeneous}
\begin{split}
\check{g} = &\frac{-i(\omega - \bm{h}\cdot \bm{\sigma})}{\sqrt{\Delta^2-( \omega - \bm{h}\cdot \bm{\sigma})^2}} \tau_3
+ \frac{\Delta}{\sqrt{\Delta^2-( \omega - \bm{h}\cdot \bm{\sigma})^2}} \tau_1.   
\end{split}
\end{equation}
In the $(\theta, \bm{M})$ parametrization, Eq.~\eqref{eq:Usadel_homogeneous} is given by
\begin{equation}
    \tan \theta = \frac{i \Delta}{\omega}, \qquad  M_0 = 1, \qquad \bm{M} = 0\,.
\end{equation}

We consider a weak perturbation of the homogeneous solution, ensuring that $\theta$ and $M_0$ remain unaffected while the triplet vector experiences perturbation.
By linearization, one can show that in the homogeneous case $\bm{M} = \frac{\bm{h}}{\Delta}$, which can be used to define the effective field $\tilde{\bm{h}} \equiv \Delta \bm{M}$ acting at a specific point.

In the inhomogeneous case, Eq.~\eqref{eq:usadel_2} becomes
\begin{equation}
\begin{split}
    \frac{\boldsymbol{h}}{\Delta} = &- \frac{1}{2} \frac{D}{\Delta} \sqrt{1-\frac{\omega^2}{\Delta^2}} \nabla^2\boldsymbol{M} + \boldsymbol{M} \Big(1 -\frac{\omega^2}{\Delta^2}\Big)\\ &+
    \frac{\sqrt{1-\omega^2/\Delta^2}}{2\Delta} \Big[\frac{1}{\tau_\mathrm{so}} -  \frac{1}{2 \tau_\mathrm{sf}}\frac{\Delta^2 + \omega^2}{\Delta^2 - \omega^2}\Big] \bm{M}\,.
\end{split}
\end{equation}
We can rearrange the equation as a screened Poisson partial differential equation by defining a screening parameter
\begin{equation}
\begin{split}
    \alpha(\omega) = &\Big(1 -\frac{\omega^2}{\Delta^2}\Big) \\ &+ \frac{\sqrt{1-\omega^2/\Delta^2}}{2\Delta} \Big[\frac{1}{\tau_\mathrm{so}} -  \frac{1}{2 \tau_\mathrm{sf}}\frac{\Delta^2 + \omega^2}{\Delta^2 - \omega^2}\Big],
\end{split}
\end{equation}

and searching for the effective field $\tilde{\bm{h}}$
\begin{equation}
    \Big[ -\frac{1}{2}\sqrt{1 - \omega^2/\Delta^2}\xi^2 \nabla^2 + \alpha(\omega) \Big] \tilde{\bm{h}} = \bm{h}\,.
\end{equation}
In two dimensions, the Green function of the homogeneous partial differential equation is
\begin{equation}
    G(\bm{r}) = \frac{1}{2\pi} K_0 (\bm{r}/ \lambda),
\end{equation}
where $\lambda = \xi  \sqrt{\frac{\sqrt{1 - \omega^2/\Delta^2}}{2 \alpha(\omega)}}$, such that the effective field is
\begin{equation}
    \tilde{\bm{h}}(\bm{r}) = \int \dd \bm{r}' G(\bm{r} - \bm{r}') \frac{\bm{h}(\bm{r}')}{\alpha(\omega)}.
\end{equation}

Therefore, the quantity $\lambda$ can be identified as the spin-averaging length.\\

\bibliography{references}

\end{document}